\documentclass[reprint,amsmath,amssymb,aps,pra]{revtex4-1}
\usepackage[utf8]{inputenc}
\usepackage{graphicx}
\usepackage{dcolumn}
\usepackage{mathrsfs,amsmath}
\usepackage{bm}
\usepackage{xcolor}
\usepackage{longtable}
\usepackage{url}

\begin{document}
\title{Global excitability and network structure in the human brain}
\author{Youssef Kora, Salma Salhi, J\"orn Davidsen, and Christoph Simon}
\affiliation{Department of Physics and Astronomy, University of Calgary, Calgary, Alberta, T2N 1N4, Canada}
\affiliation{Hotchkiss Brain Institute, University of Calgary,  Calgary,  Canada}
\date{\today}

\begin{abstract}
We utilize a model of Wilson-Cowan oscillators to investigate structure-function relationships in the human brain by means of simulations of the spontaneous dynamics of brain networks generated through human connectome data. This allows us to establish relationships between the global excitability of such networks and global structural network quantities for connectomes of two different sizes for a number of individual subjects. We compare the qualitative behavior of such correlations between biological networks and shuffled networks, the latter generated by shuffling the pairwise connectivities of the former while preserving their distribution. Our results point towards a remarkable propensity of the brain's to achieve a trade-off between low network wiring cost and strong functionality, and highlight the unique capacity of brain network topologies to exhibit a strong transition from an inactive state to a globally excited one. 
\end{abstract}
\maketitle

\section{Introduction}

The human brain is arguably the most complex and inscrutable object in the universe. Constructing a complete theory of its workings has long been considered an impracticable pursuit, but tools such as network analysis allow us to make considerable strides towards that end \cite{sporns_2003} by modeling the brain as a network of neuronal elements \cite{zalesky_fornito_bullmore_2010} to which the quantitative analysis of graph theory \cite{bullmore_2009} may be applied, and from which theoretical insights into the widely-observed neuroscientific phenomena may be obtained. Such is the general framework underlying the emerging field of network neuroscience \cite{sporns_2017}. \\ \indent
The networks employed in this sort of studies are constructed in a coarse-grained manner; the graphs typically comprise tens or hundreds of nodes, each representing a specialized, spatially segregated anatomical brain region. Those graphs in which the edges correspond to the anatomical fibre connections between the different brain regions are known as structural networks, and those in which the edges reflect statistical relationships based on similarity measures between neuronal components are referred to as functional networks \cite{lang_tome_keck_2012}. The former are typically extracted from data obtained through such non-invasive techniques as diffusion tensor imaging, and the recent advances therein have allowed for the comprehensive imaging and mapping of the structure of the human brain: the human structural connectome \cite{jbabdi_2015,shi_2017,hagmann}. Functional networks, on the other hand, are constructed by measuring  patterns of functional activity observed through such techniques as electroencephalography (EEG) and functional magnetic resonance imaging (fMRI), which shed light on the topology of both the spontaneously-emergent dynamical patterns, and those that manifest themselves during the performance of tasks \cite{eguiluz_2005, heuvel_2008, hayasaka_2010,raichle,zahra,oliver,martin}. \\ \indent
Connectomes obtained through such neuroimaging techniques, be they structural or functional, are amenable to the drawing of quantitative neuroscientific conclusions. For instance, the identification of nodes with especially strong contributions, known as hubs, may be achieved through a centrality analysis \cite{sporns2013}, in which the relative importance of different anatomical structures may be estimated by studying their betweenness within the network \cite{freeman_1977} or by looking at the eigenvector decomposition of the network \cite{bonacich_1972,bonacich_2007}. \\ \indent
Such a framework furthermore lends itself to the exploration of one of the central research areas in neuroscience, namely the study of structure-function relationships. The goal is to understand the emergence of complex neurodynamics underlain by the anatomical structure of the human brain. To that end, much of the research effort in the last two decades \cite{deco_2009,friston_2002,stephan_2004,makni_2008,sumbul_2009,hanson_2009,merlet_2013,deco_2014,deco_2015} has drawn upon the theory of dynamical systems, enabling the implementation and simulation of dynamical models constrained by the anatomical network topology \cite{motter_2006}, thereby allowing the investigation of the profound relationship between structure and function. Prominent examples of such dynamical models, that in principle allow one to generate functional connectomes from structural ones, include neural mass models \citep{david_2004,honey_2009}, oscillator models \cite{kitzbichler_2009,breakspear_2010,yan_2013,cabral_2014,odor_2019}, and spin models \cite{fraiman,marinazzo,abeyasinghe18,abeyasinghe20,abeyasinghe_2021}.
\\ \indent
We consider one such popular framework, that of Wilson-Cowan oscillators \cite{wilsoncowan}, which is a biologically motivated model for the dynamics of neuronal populations. Here, the dynamical state of the brain may be varied by the tuning of a single parameter, the parameter $c_5$; upon exceeding a certain value $c_5^T$ the system transitions from an inactive state where all activity in the network is suppressed and rapidly decays, to a globally excited state in which a multitude of brain regions exhibit rich oscillatory dynamics. This phenomenon is illustrated with an example in Fig. \ref{s111211trans}. This transition value, which is found to vary across individual subjects \cite{muldoon_2016}, may be related to a given brain network's capacity for global excitation; the lower the value of $c_5^T$, the more easily excitable the system is. 
\\ \indent
In the past, the Wilson-Cowan model has been used to reproduce the resting state fMRI activity of a simplified network with 38 cortical nodes and 2 subcortical ones \cite{deco_2009}. It has also been used to establish a connection between neurodynamics and theories of consciousness \cite{lundervold_2010}. Furthermore, in \cite{bansal_datadriven, bansal_personalized}, it was the found that global excitability, as represented by the value of $c_5^T$, predicted performance in complex cognitive tasks such as sentence completion. Motivated by these findings, we set out to explore the manner in which the human brain's propensity for global excitation is dependent on the special architecture of the the underlying networks. And so by means of implementing the Wilson-Cowan model, we simulated the resting state brain fluctuations within two types of individualized structural connectomes: the first (second) comprising 104 (84) cortical and subcortical structures, referred to henceforth as the extended (restricted) connectome. We explored the relationship between a global functional quantity such as $c_5^T$ and global structural network quantities computed for the structural connectomes, and we investigated the manner in which the transition manifested itself and how it differed between individuals. We also compared the behavior of biological networks to that of randomized networks with identical distributions of connectivities, in order to gain insight into the unique properties of the former. \\ \indent
The remainder of this paper is organized as follows: in section \ref{meth} we describe our methodology for data extraction and analysis. We present and discuss our results in section \ref{res}, and finally outline our conclusions in section \ref{conc}.

\section{Materials and Methods}\label{meth}
\subsection{Model}\label{mo}
Underlying the Wilson-Cowan model is the assumption that all neurological processes of interest are governed by the interaction between excitatory and inhibitory cells. It is furthermore assumed that each subpopulation of such cells at every brain region may be characterized by a single variable. Defining $E_i$ and $I_i$ as the respective fractions of excitatory and inhibitory subpopulations of region $i$, the Wilson-Cowan model reads 
\begin{multline}\label{wilcowe}
\tau \frac{dE_i}{dt} = -E_i(t)+(S_{E_m}-E_i(t)) \\ 
S_E \left( c_1E_i(t)-c_2I_i(t) +c_5\sum_j J_{ij}E_j(t-\tau_d^{ij})+P_i(t) \right) \\
+ \sigma w_i(t)
\end{multline}
and
\begin{multline}\label{wilcowr}
\tau \frac{dI_i}{dt} = -I_i(t)+(S_{I_m}-I_i(t)) \\ 
S_I \left( c_3E_i(t)-c_4I_i(t) +c_6\sum_j J_{ij}I_j(t-\tau_d^{ij}) \right) \\
+ \sigma v_i(t)
\end{multline}
where $S_{E,I}(x)$ are sigmoid functions given by
\begin{eqnarray}\label{sig}
S_{E,I}(x) = \frac{1}{1+e^{-a_{E,I}(x-\theta_{E,I})}}
- \frac{1}{1+e^{a_{E,I}\theta_{E,I}}}
\end{eqnarray}
and $S_{E_m,I_m}$ are the maxima thereof, and the constants $a_{E,I}$ and $\theta_{E,I}$ respectively determine the value and position of maximum slope. $J_{ij}$ refers to the elements of the anatomical connectivity matrix. An external stimulation $P_i(t)$ may be applied to excitatory cells. The finite distance $d_{ij}$ between two brain regions gives rise to the communication delay $\tau_d^{ij}=d_{ij}/v_d$, where $v_d=10 m/s$ is the signal transmission velocity. A normal distribution of noise is added to the system through the functions $w_i(t)$ and $v_i(t)$, with strength $\sigma$. 
\\ \indent
An intuitive understanding of the model may be acquired by considering the second term on the right hand sides of both equations, which represents the respective activity in each subpopulation, i.e., the proportion of cells which meet the two conditions of a) being sensitive to an excitation (not in a refractory state) and b) receiving at least threshold excitation at the same time $t$. The probability for the former condition is represented by the bracket multiplying the sigmoid function, and the probability for the latter condition is represented by the sigmoid function itself. The choice of the sigmoid function is fundamentally based on empirical observations of both single cell and population response curves \cite{daniel1965,rall1965}, but the underlying intuition is straightforward: too small an excitation will fail to excite any elements at all, whereas a very large one can do no more than excite the entirety of the population. The argument of the sigmoid function is, of course, the excitation itself, which is enhanced by excitatory cells, suppressed by inhibitory ones, contributed towards by the other nodes through the structural connectivity matrix, and potentially by a stimulation $P_i(t)$.
\\ \indent
It is standard to use biologically derived values for all the parameters in the model besides $c_5$ and $c_6$, which respectively represent the excitatory and inhibitory global coupling strengths between all brain regions in the network. As in \cite{bansal_datadriven}, we set $c_6=c_5/4$ based on an approximate ratio between excitatory and inhibitory coupling. Thus $c_5$ remains the only free parameter, and its tuning, as mentioned above, determines the dynamical state of the brain in the model. \\ \indent
\subsection{Structural data}\label{strdata}
We constrained the Wilson-Cowan dynamical model by the structural connectivity data produced from the 1200 subject cohort of the Human Connectome Project (HCP)\href{http://www.humanconnectomeproject.org/}, a database containing neural data for thousands of subjects, of which we selected a sample for our calculations. Both pre-processed T1-weighted structural images and 3T dMRI images were used in our computational fibre tracking method. Python DIPY and NiBabel libraries were utilized to perform the streamline calculations using a constrained spherical deconvolution model and probabilistic fibre tracking functions, which are built in the libraries. 
\\ \indent
The product of this procedure was a set of structural connectivity networks, each belonging to an individual subject and comprising 104 nodes corresponding to cortical or subcortical brain structures, the full list of which may be found in Table \ref{tab:extended_list} in the Appendix. The networks were then normalized by dividing the strength of each connection by the sum of the
volumes of its two nodes, as in earlier studies \cite{muldoon_2016,bansal_datadriven} (for a comparison with results for non-normalized connectomes, see Fig. \ref{allnon} in the Appendix.). The matrices representing the physical distances between brain regions were obtained by using the mapping algorithm in our code to determine the coordinates of the nodes and then computing the distances between them, which, in so coarse-grained a parcellation of the human brain, were simply approximated as Euclidean distances. 
\\ \indent
In an effort to investigate the effect of the atlas utilized, we also generated restricted connectomes for the same set of subjects, this time with 84 brain regions. The structures excluded from the restricted connectomes include the brainstem among a number of other subcortical regions, the full list of which  may be found in Table \ref{tab:removed_structures} in the Appendix. These 84 brain regions constitute the FreeSurfer Desikan-Killiany atlas, which is often used in studies relying on the MRTrix software (see \cite{salhi_2022} for more details).  \\ \indent
Finally, we generated for each subject a network with shuffled connectivities, while preserving the overall distribution of connectivies (but not the distribution of degree centralities). This is accomplished by randomly rearranging the entries of the original 104x104 matrix ${\bf J}$ below the diagonal, while preserving their numerical values, and then reflecting them about the diagonal to restore the symmetry of the matrix. This allowed us to study the way in which structure-function relationships are affected by the specific arrangement of anatomical connectivities in the brain. To distinguish these shuffled connectomes from the ones obtained from imaging data, we shall henceforth refer to the latter as biological connectomes, be they extended (comprising 104 brain regions) or restricted (comprising 84 brain regions)
\subsection{Network Characteristics}\label{netchar}
A number of global structural properties were computed for each network. The first of these is the characteristic path length, which is a global measure of how strongly connected a network is \cite{stam_2007}. It is computed as average path length between all possible pairs of vertices:
\begin{eqnarray}\label{path}
L = \frac{1}{N(N-1)}\sum_{i,j\in N, i \neq j} s_{ij}
\end{eqnarray}
Where $N$ is the number of nodes, and $s_{ij}$ is the shortest distance between nodes $i$ and $j$ (note that this graph-theoretical distance $s_{ij}$ is not to be confused with the physical distance $d_{ij}$). In the case of weighted graphs such as our own, we computed the distance between two nodes as simply the reciprocal of the strength of the connection between them. i.e., $1/J_{ij}$, and the Dijkstra shortest path \cite{dijkstra} $s_{ij}$ was then computed for every pair of nodes. Another closely related measure is the average degree, i.e., the total connectivity of a given node averaged across all nodes, given by
\begin{eqnarray}\label{degcent}
K = \frac{1}{N(N-1)} \sum_{i,j \in N}  J_{ij}
\end{eqnarray}
We also computed the spectral radius $R$, which is related to the variability of the degrees of nodes \cite{meghanathan}, and is simply given by the largest eigenvalue of the connectivity matrix. Finally, we computed the synchronizability of the networks, given by $\frac{\lambda_2^L}{\lambda_{max}^L}$, with $\lambda_2^L$ and $\lambda_{max}^L$ respectively being the second-to-smallest and the largest eigenvalue of the Laplacian matrix $L=D-J$, which is the difference between the degree matrix and the connectivity matrix  \cite{boccaletti}. This ratio has been shown to determine synchronizability for oscillator networks by assessing the linear stability of the synchronous state \cite{barahona_2002}. We then measured the correlations of all these structural quantities with the functional quantitiy $c_5^T$ in an effort to gain insight into structure-function relationships within this framework. \\ \indent
Finally, we quantify the degree of hierarchical structure within networks by means of the global reaching centrality (GRC) \cite{mones_2012}, for which computation we employed the Python package NetworkX \cite{networkx}. GRC is defined, for a graph $G$ with $N$ nodes, as
\begin{eqnarray}\label{globr}
GRC = \frac{\sum_{i \in V}[C_R^{max}-C_R(i)]}{N-1}
\end{eqnarray}
where $V$ is the set of nodes in $G$, $C_R(i)$ is the local reaching centrality of node $i$, and $C_R^{max}$ is the maximum value thereof. Local reaching centrality is defined for unweighted directed graphs as the fraction of all nodes in the network that may be reached from node $i$, and generalized for weighted graphs as
\begin{eqnarray}\label{locr}
C_R(i) = \frac{1}{N-1} \sum_{j:0<n^{out}(i,j)<\infty} \frac{\sum_{k=1}^{n^{out}(i,j)}w_i^{(k)}(j)}{n^{out}(i,j)}
\end{eqnarray}
with $n^{out}(i,j)$ being the number of links along the shortest path from node $i$ to node $j$, and $w_i^{(k)}(j)$ is the weight of the $k$-th such link. 

\begin{figure}[h]
\centering
\includegraphics[width=0.47\textwidth]{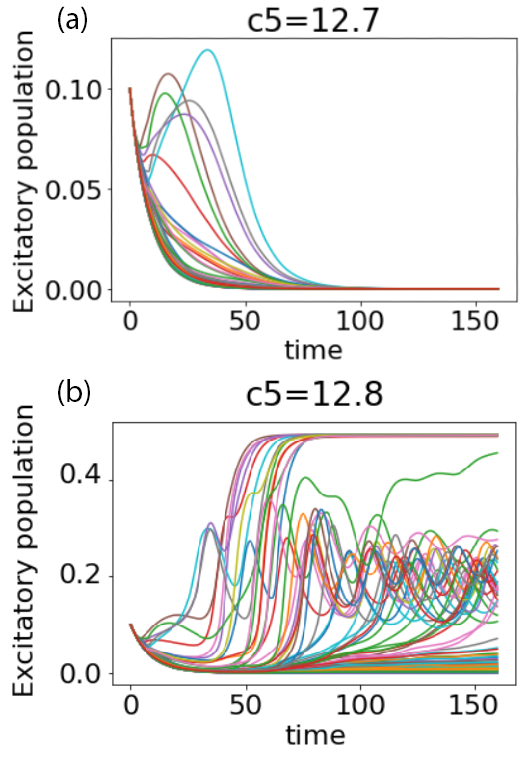}
\caption{The dynamics of the excitatory population of subject 111211 at $c_5=12.7$ (a) and $c_5=12.8$ (b). Each line corresponds to a brain region.}
\label{s111211trans}     
\end{figure}

\subsection{Simulation Details}\label{simdet}
The dynamics were simulated using the Wilson-Cowan model, as explained above, in the absence of an external stimulation. We utilized a second order Runge-Kutta solver with a sufficiently fine timestep, such that the results were independent of the size thereof. For each individual subject, the value of $c_5^T$ was estimated by simulating the model at multiple values of $c_5$ and observing the point at which the transition took place.
\\ \indent
We switched off the external stimulation, initialized all oscillators at $E=I=0.1$, and chose the parameters $\sigma=10^{-5}$, $c_1=16$, $c_2=12$, $c_3=15$, $c_4=3$, $a_E=1.3$, $a_I=2$, $\theta_E=4$, $\theta_I=3.7$, and $\tau=8$ as prescribed in the literature \cite{muldoon_2016,bansal_datadriven}. At these values of the parameters, the oscillators can be in one of three states: a low fixed point, a high fixed point, and an oscillatory limit cycle inbetween \cite{muldoon_2016}. For every subject, we observed the strong transition (as discussed in the introduction) into a globally excited state, i.e., from a state where activity in all oscillators decays into the low fixed point, to a state where a significant fraction of the oscillators transition into either the limit cycle or the high fixed point. As mentioned above, this is achieved by varying the global coupling parameter $c_5$, and the value at which this transition takes place in the absence of an external stimulation, $c_5^T$, is unique for each subject, at a given choice of parameters and initial conditions. We demonstrate the effect in Fig. \ref{s111211trans}, which displays the fractional excitatory population in each brain region as a function of time for a given subject, and in the absence of an external stimulation. \\ \indent
\section{Results and Discussion}\label{res}
\begin{figure}[h]
\centering
\includegraphics[width=0.47\textwidth]{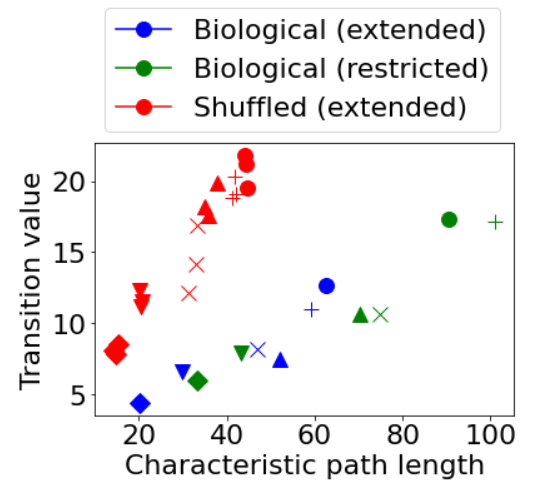}
    \caption{The transition value $c_5^T$ vs. the characteristic path length of a network. Symbol shapes correspond to individual subjects. Blue symbols are extended connectomes, green are restricted connectomes (in accordance with the FreeSurfer Desikan-Killiany atlas), and red are extended connectomes with randomly redistributed connectivities (preserving the overall distribution of connectivities). Errors are smaller than the symbol sizes.}
\label{all}     
\end{figure}
We start by comparing all three types of connectomes (extended, restricted, and shuffled) in Fig. \ref{all}, in which we plot for each type of connectome the transition value $c_5^T$ vs. the characteristic path length $L$. 
Each of the 6 different symbols corresponds to a different subject.  A few things may be noted by observing Fig. \ref{all}. Firstly, both sorts of biological networks fall on the same trendline ($r^2=0.879$). This is a simple observation that more strongly connected networks have a higher tendency for global excitation. \\ \indent
Secondly, by comparing the green points (corresponding to the FreeSurfer Desikan-Killainy atlas) to the blue points in Fig. \ref{all}, one observes that the restricted network for any given subject invariably has a higher characteristic path length and higher transition value than its extended counterpart. This suggests that the exclusion of the brainstem (and the other 19 structures absent from the FreeSurfer Desikan-Killiany atlas) causes the networks to be substantially less well-connected, and thus harder to globally excite. This observation is consistent with the findings in \cite{salhi_2022} highlighting the structural importance of the brainstem, and of incorporating the full list of subcortical structures. We eliminated the possibility that this difference is an artifact of the system size by performing the same calculations for another set of restricted connectomes with 20 excluded structures, but now with the latter's being randomly chosen. The list of such structures, the same for every subject, may be found in table \ref{tab:removed_structures_2} in the Appendix, and the result of the analysis may be found in Fig. \ref{allrr} and the associated text, also in the Appendix. \\ \indent
The third observation is perhaps the most striking. It is clear that shuffling the connectivities of extended networks (as represented by the red points in Fig. \ref{all}) invariably shortens the characteristic path length while causing the transition value to rise, which causes the red points to deviate from the biological trendline. Instead, the random networks form their own displaced trendline. We find it rather curious, not only that a biological network should always be less strongly-connected compared to a random network with the same distribution of connectivities, but that it should simultaneously have a higher global excitability. This is the reverse of the phenomenon observed upon excluding the 20 subcortical structures (going from the blue points to the green points), which caused the networks to both be less well-connected and less globally excitable.  \\ \indent
The tendency of random networks to have a low characteristic path length is by itself an unsurprising finding, as random networks tend to have a low mean path length \cite{stam_2007,bullmore_2009}. However, that global excitability should fall in spite of that is quite an interesting observation and has implications for the economics of brain network organization. The creation and maintenance of anatomical connections is an expensive affair, and the brain must therefore sagaciously utilize resources within the network in order to bring about the desired topological properties in an efficient way \cite{bullmore_2012}. Consistently with this trade-off principle, the results in Fig. \ref{all} suggest that brain networks are wired in a manner that is both parsimonious and uncompromising of the required topology for high global excitability. \\ \indent
\begin{figure}[h]
\centering
\includegraphics[width=0.47\textwidth]{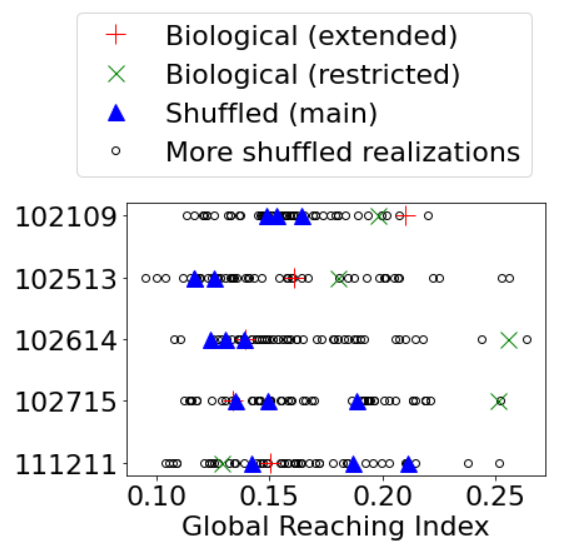}
    \caption{The global reaching centrality of the biological extended (blue triangles), biological restricted (green X's), the 3 main realizations of shuffled networks (red pluses), and 50 more realizations of the latter (black hollow circles). The results are presented for 5 subjects.}
\label{grc}     
\end{figure}
As a possible underlying cause for this disparity in behavior between biological and shuffled networks, one candidate is the hierarchical structure of biological networks \cite{meunier_2010,park_2013}. In particular, it has been demonstrated that the hierarchical manner in which connectivity is distributed within the human connectome is responsible for giving rise to critical dynamics over an extended region of parameter space, known as the Griffiths phase, rather than a singular critical point (see, for instance, Refs.~\cite{moretti_2013,odor_2015,li_2017}). There are a number of ways to quantify the extent of hierarchy within a network~\cite{mones_2012,murtra_2013}, which allow us to test the hypothesis that these hierarchies are largely preserved in going from the biological extended connectome to the biological restricted one, but are destroyed when pairwise connectivities are randomly shuffled. \\ \indent
\begin{figure}[h]
\centering
\includegraphics[width=0.36\textwidth]{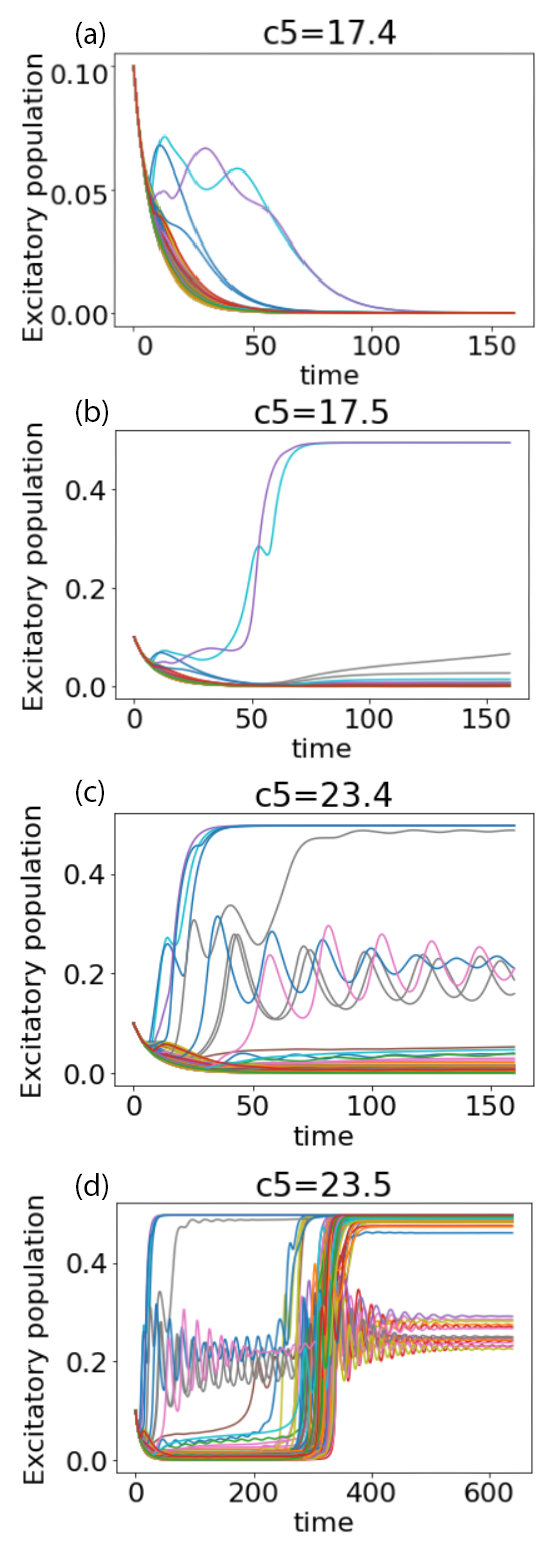}
\caption{The dynamics of the excitatory populations of the shuffled network of subject 111211, at $c_5=17.4$ (a), $c_5=17.5$ (b), $c_5=23.4$ (c), and 23.5 (d). Here, $c_5^T=23.4$.}
\label{shuffled}     
\end{figure}
Specifically, to investigate this hypothesis, we computed the global reaching centrality (GRC) (defined in subsection \ref{netchar}) for the biological extended connectome, the restricted one, the 3 realizations of shuffling presented earlier, and 50 additional realizations of shuffling.  The results, presented in Fig. \ref{grc} for 5 different subjects, suggest that the shuffling process does not in general tend to degrade the hierarchical structure of the biological connectomes, at least not within our rather coarse-grained connectomes. Thus, our GRC analysis indicates that the high global excitability in our human brain connectomes is not due to any hierarchical architecture. \\ \indent
\begin{figure}[h]
\centering
\includegraphics[width=0.42\textwidth]{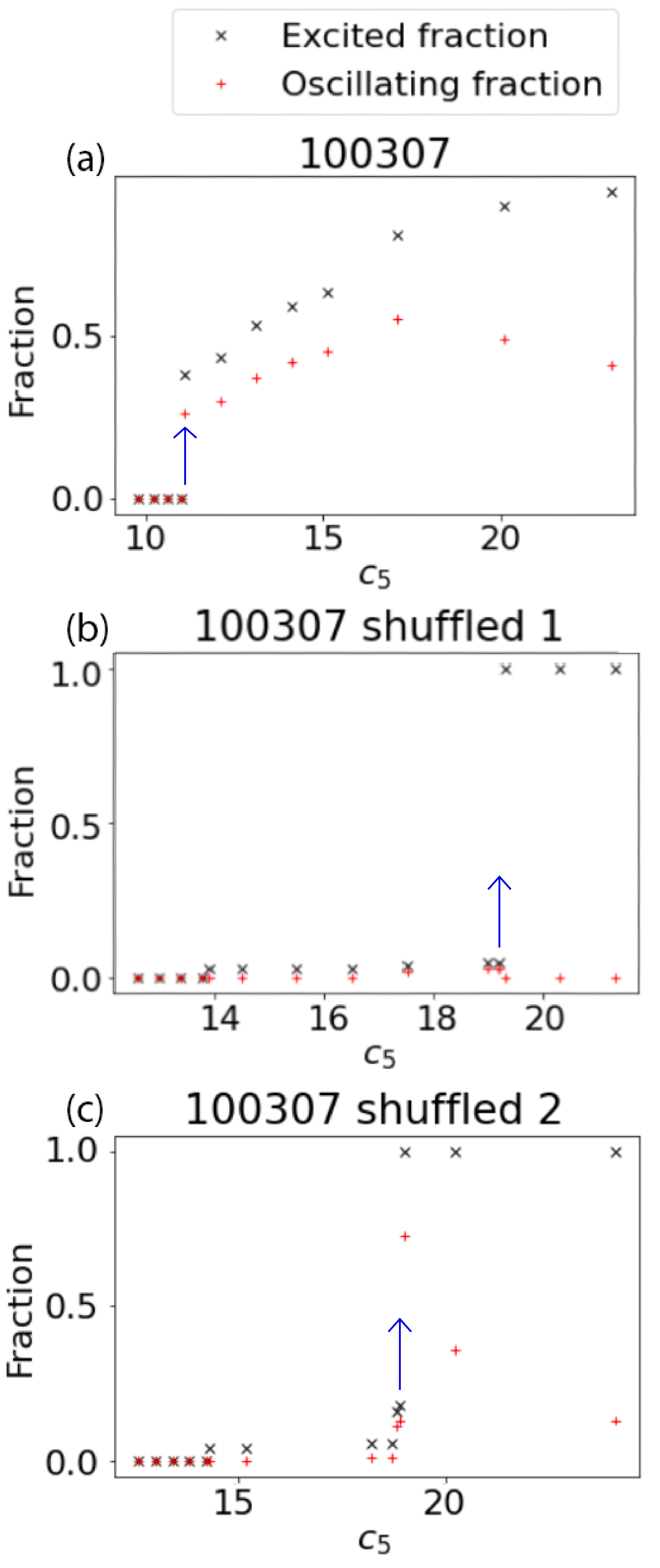}
\caption{The excited fraction (crosses) and oscillating fraction (circles) of brain regions for the biological network (a) and two instances of shuffling for that same network (b \& c) belonging to subject 100307. Blue arrows indicate the location of the transition.}
\label{osc100307}     
\end{figure}
\begin{figure}[h]
\centering
\includegraphics[width=0.47\textwidth]{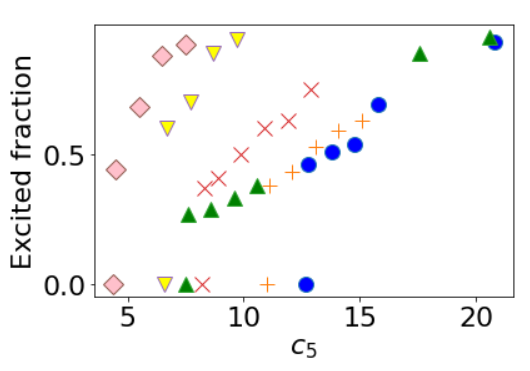}
    \caption{The excited fraction as a function of $c_5$. Symbol shapes and colors correspond to subject names. Errors are small than symbol sizes.}
\label{exfrac}     
\end{figure}
Another notable difference between biological and shuffled networks is in the manner in which the transition manifests itself. This becomes clear by comparing Fig. \ref{s111211trans} and Fig \ref{shuffled}. In the former case, there is a sudden, rather dramatic increase of activity upon exceeding the transition value. In the latter case of shuffled networks, this is not always true. Instead, two transitions might be present: At low values of $c_5$ a transition occurs to a state where only a very small fraction of brain regions are excited, and then at a sufficiently high value of $c_5$ an abrupt jump in activity occurs, similar to the biological networks, which is, hence, denoted as $c_5^T$ (for an analysis with $c_5^T$ alternatively defined, see Fig. \ref{allalt} and associated text in the Appendix). \\ \indent
This phenomenon may also be observed in Fig. \ref{osc100307}, which presents the evolution of the excited fraction (the total fraction of brain regions that saturate at the high fixed point or the limit cycle) and the oscillatory fraction (the fraction that includes only those brain regions that saturate at the limit cycle). In all cases, the excited fraction starts at zero at low values of $c_5$, and saturates to unity at sufficiently high values of $c_5$. The differences, however, is in the manner in which the excited fraction evolves from zero to unity. In biological networks, we invariably observed a well-defined jump from zero to a finite value of the excited fraction at a specific value of $c_5$, to which we refer as $c_5^T$, and beyond which the excited fraction proceeds to smoothly grow until it saturates at unity. A typical example of this behavior is presented in Fig. \ref{osc100307}a. The situation in shuffled networks is quite different, however; the excited fraction initially starts at zero, and as we raise the value of $c_5$, the system might leave the unexcited ground state at a certain value of $c_5$ but manifesting excitations in only a very small fraction of brain regions, until a sufficiently high value of $c_5$ (which, as previously mentioned, we define as $c_5^T$) is reached, whereupon {\em all} brain regions become excited, and remain as such as we further increase $c_5$. This region of meager activity that precedes the transition was often (but not always) observed in shuffled networks, and it was never seen in any of the biological networks examined. The jump at $c_5^T$, however, which takes the system from zero (or near zero) to unity was observed in all realizations of shuffled networks. This behavior may be observed in Fig. \ref{osc100307}b and Fig. \ref{osc100307}c, which present the behavior for two instances of shuffling for the same network. \\ \indent
The behavior of shuffled networks is, however, quite variable. For example, the two realizations of shuffling presented in Fig. \ref{osc100307}b and Fig. \ref{osc100307}c, respectively, show clear differences in the behavior of the oscillatory fraction. It is clear that in the case of the shuffled network in Fig. \ref{osc100307}b, the oscillatory fraction never gains a significant value at any point, even after the transition takes place at $c_5^T=19.2$. This is in clear contrast with the case in Fig. \ref{osc100307}c, in which a significant fraction of brain regions exhibit non-trivial oscillations beyond the transition, albeit rapidly decaying as we further raise the value of $c_5$. These qualitatively different classes of behavior suggest that the specific way in which anatomical connections are arranged in the brain is crucial for guaranteeing a strong transition into a state with prolific oscillatory dynamics as exhibited in Fig. \ref{osc100307}a.
\\ \indent
\begin{figure}[h]
\centering
\includegraphics[width=0.47\textwidth]{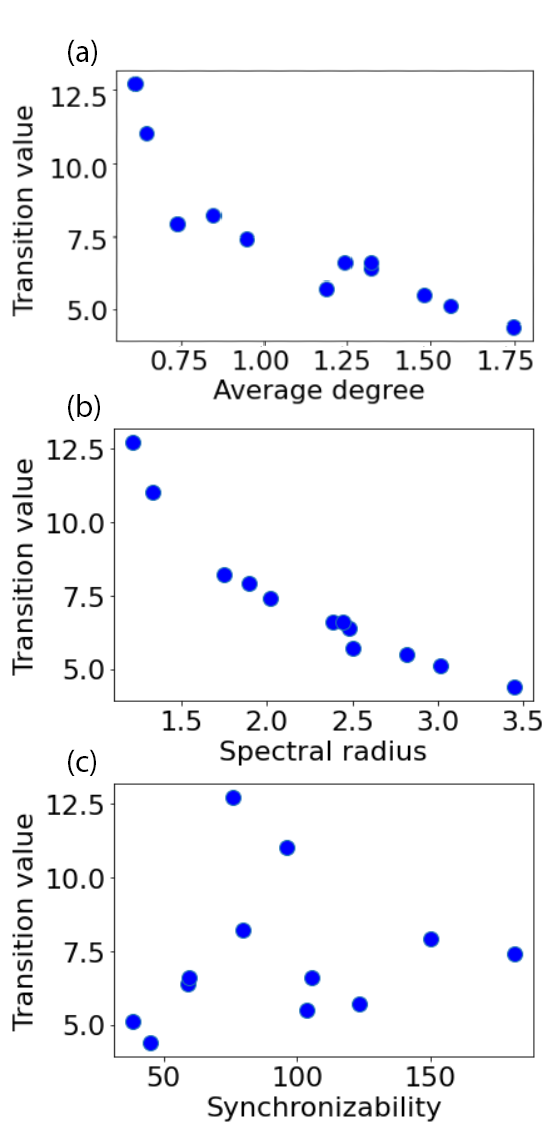}
\caption{The transition value $c_5^T$ vs. the average degree (a), spectral radius (b), and synchronizability (c), for the 6 subjects previously considered, in addition to 6 others. The values of $r^2$ are respectively 0.796, 0.885, and 0.0262.}
\label{corrs}     
\end{figure}
We also conducted a closer inspection of the manner in which the transition manifests itself in the case of biological networks, and how it varies across individuals. In Fig. \ref{exfrac}, we plot the excited fraction as a function of $c_5$ for each of the six subjects considered in this study. Every biological network experienced a considerable jump in the excited fraction upon crossing its respective value of $c_5^T$, as mentioned above. But not only was the size of the jump different across subjects and seemingly uncorrelated with the value of $c_5^T$, but so was the rate at which the excited fraction continued to grow. This demonstrates that individual variability across subjects is not limited to the value of $c_5^T$ itself, but that the character of the states beyond the transition, as viewed by studying the growth of the excited fraction, is another independent consideration. \\ \indent
Finally, we measured the correlation between $c_5^T$ and a variety of structural quantities (defined in subsection \ref{netchar}), and present them in Fig. \ref{corrs}. The average degree and spectral radius are both measures of the overall connectivity of the network, and thus it was not unexpected that they should negatively correlate with the transition value ($r^2= 0.796,0.885$ respectively), in light of the result presented in Fig. \ref{all}. The difference, however, between the average degree and the spectral radius is that the former is invariant under the reshuffling of connectivities while the latter is not. Lastly, there was no correlation to be found with global synchronizability ($r^2=0.0899$). The latter observation, which suggests that path length and synchronizability are not directly related concepts, is consistent with the `paradox of heterogeneity', which is the observation that (unweighted and undirected) graphs with homogeneous connectivity distributions tend to synchronize more readily than those with heterogeneous ones, irrespective of the latter's possessing lower average path lengths \cite{stam_2007,nishikawa}. \\ \indent
 A possible extension of this work might be to attempt to connect these results to the critical brain hypothesis, which states that the brain self-tunes to a regime with a large dynamical range that is reminiscent of a critical point or regime in statistical physics~\cite{kinouchi_2006,moretti_2013}. Indeed, using the Ising model at its critical point a recent study has calculated the dimensionality of the brain and found it to vary between healthy groups and those with disorders of consciousness \cite{abeyasinghe20}. More fundamentally, the critical brain hypothesis is partially based on the observation of scale-free information or spreading cascades, so-called neuronal avalanches, across species and spatial scales~\cite{cocchi_2017,yaghoubi_2018,korchinski_2021, curic_2021}. Recent theoretical work suggests that scale-free neuronal avalanches require an "edge-of-synchronization" phase transition~\cite{disanto_2018}. Our work here further suggests that $c_5^T$ could serve as this phase transition point but more work is needed to explore this in the future. \\ \indent
On that note, our results could also be examined within the context of popular theories of consciousness. Global workspace theory (GWT) is one such theory which suggests that the brain has a fleeting memory capacity, enabling back and forth access between separate brain functions \cite{baars_2005}. Specifically, this theory purports that the global network is activated non-linearly, through a process known as ``ignition" \cite{dehaene_sergent_changeux_2003}, which is the sudden activation of neurons that code for current conscious content \cite{mashour_roelfsema_changeux_dehaene_2020}. It has been shown that a reduction in the interconnectivity of global network neurons makes ignition more difficult to reach \cite{mashour_roelfsema_changeux_dehaene_2020}, suggesting that GWT requires a network that is optimized for high excitability. Another popular theory is integrated information theory (IIT), which identifies and attempts to quantify the capacity of a network to integrate information \citep{zhao_zhu_tang_xie_zhu_zhang_2019}, represented by the quantity $\Phi$, often referred to as the quantity of conscious experience. In \cite{popiel}, $\Phi$ was computed for small Ising systems and was found to exhibit criticality at the temperature of the Ising phase transition.  While different in their approach, both theories require an efficient functional network. Our results, which suggest that biological networks are uniquely designed to possess great global excitability and strongly transition from an inactive state to a state with rich oscillatory dynamics, could be used to further analyze these theories and shed more light on the mechanism of consciousness.

\section{Conclusions}\label{conc}
We studied the spontaneous functional activity that arises in 6 brain networks, each belonging to a different subject, by using the Wilson-Cowan model to simulate the network dynamics in the absence of an external stimulation. Under this framework, the dynamical state of the brain is dictated by the numerical value of a single parameter $c_5$, and at a critical value thereof, referred to as $c_5^T$, these biological networks exhibit a strong transition from a state where no activity is allowed to propagate, to a globally excited state with rich oscillatory dynamics spanning a significant fraction of brain regions. Our results confirm that the value of $c_5^T$ displays significant variability across individuals, and further show that this variability extends to the character of the transition as observed from the behavior of the excited fraction as a function of $c_5$. We have also drawn insights into the remarkable ability of the brain to comprise networks designed in a manner that is both resource-efficient and conducive to the desirable functional properties: our observations suggest that biological networks form their connectivities so judiciously as to give rise to high global excitability while simultaneously attempting to keep the associated wiring cost reasonably low. Furthermore, it was clear that an invariably strong transition from a globally inactive to one with ubiquitous oscillatory dynamics is a special property of biological networks, and is in general not enjoyed by their reshuffled counterparts. We also established correlations between $c_5^T$ and a number of network properties such characteristic path length, average degree, and spectral radius, but found it to be uncorrelated with network synchronizability. Finally, our results indicate that utilizing a restricted atlas of the human brain causes the networks to become both less well-connected and less susceptible to global excitation, in a manner consistent with the biological relationship we established between $c_5^T$ and the characteristic path length.

\section{Acknowledgments}\label{ack}
This work was supported by the Natural Sciences and Engineering Research Council (NSERC) of Canada and the Alberta Major Innovation Fund. We would also like to thank Wilten Nicola, Davor Curic, and Omid Khajehdehi from the Complexity Science Group (CSG) for the discussion and input. Additionally, we thank the HCP for providing access to their data.

\section*{Data Availability Statement}
The data used in this project was provided by the Human Connectome Project (HCP; Principal
Investigators: Bruce Rosen, M.D., Ph.D., Arthur W. Toga, Ph.D., Van J. Weeden, MD). HCP funding was provided by the
National Institute of Dental and Craniofacial Research (NIDCR), the National Institute of Mental Health (NIMH), and the
National Institute of Neurological Disorders and Stroke (NINDS). HCP data are disseminated by the Laboratory of Neuro
Imaging at the University of Southern California. Structural and diffusion MRI images from the HCP, as well as lists of extracted structures, bvals, and bvecs, were all used to process the data in our Python program. All subjects are part of the ``WU-Minn HCP Data - 1200 Subjects" \href{https://db.humanconnectome.org/data/projects/HCP_1200}{dataset}. A complete list of subject names is available upon request. All scripts used to generate the connectomes are available on our \href{https://github.com/SalmaSalhi7/Structural-Connectome-Project}{Github} repository found at \url{https://github.com/SalmaSalhi7/Structural-Connectome-Project}.

\bibliography{biblio}

\section{Appendix}
\subsection{Lists of Brain regions}

 \begin{longtable}{|c | c |  c| c| c|} 
  \caption{A full list of the 104 brain structures.}
\\ \hline
ID & Structure  
\\ \hline
1 & Left-Lateral-Ventricle\\ 
2 & Left-Inf-Lat-Vent  \\ 
3 & Left-Cerebellum-Cortex  \\
4 & Left-Thalamus-Proper \\ 
        5 & Left-Caudate \\ 
        6 & Left-Putamen  \\ 
        7 & Left-Pallidum \\ 
        8 & 3rd-Ventricle  \\ 
        9 & 4th-Ventricle  \\ 
        10 & Brain-Stem  \\ 
        11 & Left-Hippocampus  \\ 
        12 & Left-Amygdala  \\ 
        13 & CSF  \\ 
        14 & Left-Accumbens-area  \\ 
        15 & Left-VentralDC \\ 
        16 & Left-vessel  \\ 
        17 & Left-choroid-plexus \\ 
        18 & Right-Lateral-Ventricle  \\ 
        19 & Right-Inf-Lat-Vent  \\ 
        20 & Right-Cerebellum-Cortex  \\ 
        21 & Right-Thalamus-Proper  \\ 
        22 & Right-Caudate  \\ 
        23 & Right-Putamen \\ 
        24 & Right-Pallidum \\ 
        25 & Right-Hippocampus  \\ 
        26 & Right-Amygdala \\ 
        27 & Right-Accumbens-area  \\ 
        28 & Right-VentralDC \\ 
        29 & Right-vessel  \\ 
        30 & Right-choroid-plexus  \\ 
        31 & Optic-Chiasm  \\ 
        32 & CC\_Posterior \\ 
        33 & CC\_Mid\_Posterior \\ 
        34 & CC\_Central \\ 
        35 & CC\_Mid\_Anterior \\ 
        36 & CC\_Anterior  \\ 
        37 & ctx-lh-bankssts \\ 
        38 & ctx-lh-caudalanteriorcingulate \\ 
        39 & ctx-lh-caudalmiddlefrontal  \\ 
        40 & ctx-lh-cuneus  \\ 
        41 & ctx-lh-entorhinal  \\ 
        42 & ctx-lh-fusiform  \\ 
        43 & ctx-lh-inferiorparietal  \\ 
        44 & ctx-lh-inferiortemporal  \\ 
        45 & ctx-lh-isthmuscingulate  \\ 
        46 & ctx-lh-lateraloccipital \\ 
        47 & ctx-lh-lateralorbitofrontal  \\ 
        48 & ctx-lh-lingual \\ 
        49 & ctx-lh-medialorbitofrontal \\ 
        50 & ctx-lh-middletemporal  \\ 
        51 & ctx-lh-parahippocampal  \\ 
        52 & ctx-lh-paracentral  \\ 
        53 & ctx-lh-parsopercularis \\ 
        54 & ctx-lh-parsorbitalis  \\ 
        55 & ctx-lh-parstriangularis \\ 
        56 & ctx-lh-pericalcarine  \\ 
        57 & ctx-lh-postcentral \\ 
        58 & ctx-lh-posteriorcingulate \\ 
        59 & ctx-lh-precentral  \\ 
        60 & ctx-lh-precuneus  \\ 
        61 & ctx-lh-rostralanteriorcingulate \\ 
        62 & ctx-lh-rostralmiddlefrontal \\ 
        63 & ctx-lh-superiorfrontal  \\ 
        64 & ctx-lh-superiorparietal \\ 
        65 & ctx-lh-superiortemporal  \\ 
        66 & ctx-lh-supramarginal  \\ 
        67 & ctx-lh-frontalpole  \\ 
        68 & ctx-lh-temporalpole \\ 
        69 & ctx-lh-transversetemporal \\ 
        70 & ctx-lh-insula \\ 
        71 & ctx-rh-bankssts  \\ 
        72 & ctx-rh-caudalanteriorcingulate \\ 
        73 & ctx-rh-caudalmiddlefrontal  \\ 
        74 & ctx-rh-cuneus \\ 
        75 & ctx-rh-entorhinal \\ 
        76 & ctx-rh-fusiform  \\ 
        77 & ctx-rh-inferiorparietal  \\ 
        78 & ctx-rh-inferiortemporal \\ 
        79 & ctx-rh-isthmuscingulate  \\ 
        80 & ctx-rh-lateraloccipital  \\ 
        81 & ctx-rh-lateralorbitofrontal \\ 
        82 & ctx-rh-lingual  \\ 
        83 & ctx-rh-medialorbitofrontal  \\ 
        84 & ctx-rh-middletemporal  \\ 
        85 & ctx-rh-parahippocampal\\ 
        86 & ctx-rh-paracentral \\ 
        87 & ctx-rh-parsopercularis \\ 
        88 & ctx-rh-parsorbitalis \\ 
        89 & ctx-rh-parstriangularis \\ 
        90 & ctx-rh-pericalcarine  \\ 
        91 & ctx-rh-postcentral \\ 
        92 & ctx-rh-posteriorcingulate  \\ 
        93 & ctx-rh-precentral  \\ 
        94 & ctx-rh-precuneus  \\ 
        95 & ctx-rh-rostralanteriorcingulate \\ 
        96 & ctx-rh-rostralmiddlefrontal  \\ 
        97 & ctx-rh-superiorfrontal \\ 
        98 & ctx-rh-superiorparietal  \\ 
        99 & ctx-rh-superiortemporal \\
        100 & ctx-rh-supramarginal  \\ 
        101 & ctx-rh-frontalpole  \\ 
        102 & ctx-rh-temporalpole \\ 
        103 & ctx-rh-transversetemporal  \\ 
        104 & ctx-rh-insula  \\ [1ex] 
 \hline 
 \end{longtable}
        \label{tab:extended_list}

\begin{table}[t]
 \caption{The list of 20 structures removed in the restricted connectome corresponding to the FreeSurfer Desikan-Killiany atlas.}
    \centering
    \begin{tabular}{|c|}
        \hline
        Removed Structures \\ \hline 
        Left-Lateral-Ventricle \\
        Left-Inf-Lat-Vent \\
        3rd-Ventricle \\
        4th-Ventricle \\
        Brain-Stem \\
        CSF \\
        Left-VentralDC \\
        Left-vessel \\
        Left-choroid-plexus \\
        Right-Lateral-Ventricle \\
        Right-Inf-Lat-Vent \\
        Right-VentralDC \\
        Right-vessel \\
        Right-choroid-plexus \\
        Optic-Chiasm \\
        CC\_Posterior \\
        CC\_Mid\_Posterior \\
        CC\_Central \\
        CC\_Mid\_Anterior \\
        CC\_Anterior \\ \hline
    \end{tabular}      
    \label{tab:removed_structures}
\end{table}

\begin{table}[t]
   \caption{The list of 20 structures removed in the randomly restricted connectome.}
    \centering
    \begin{tabular}{|c|}
        \hline
        Removed Structures \\ \hline 
        ctx-lh-middletemporal \\
        ctx-lh-caudalmiddlefrontal \\
        ctx-rh-pericalcarine \\
        ctx-rh-bankssts \\
        Left-VentralDC \\
        ctx-lh-transversetemporal \\
        ctx-lh-cuneus \\
        Left-Inf-Lat-Vent \\
        ctx-lh-frontalpole \\
        ctx-lh-paracentral \\
        Optic-Chiasm  \\
        ctx-lh-fusiform  \\
        Left-Amygdala \\
        ctx-rh-transversetemporal \\
        ctx-lh-precuneus \\
        ctx-lh-rostralanteriorcingulate \\
        Left-Pallidum  \\
        ctx-rh-inferiorparietal  \\
        ctx-rh-rostralanteriorcingulate \\
        ctx-lh-insula \\ \hline
    \end{tabular}      
    \label{tab:removed_structures_2}
\end{table}

\subsection{Supplementary Figures}
\begin{figure}[h]
\centering
\includegraphics[width=0.47\textwidth]{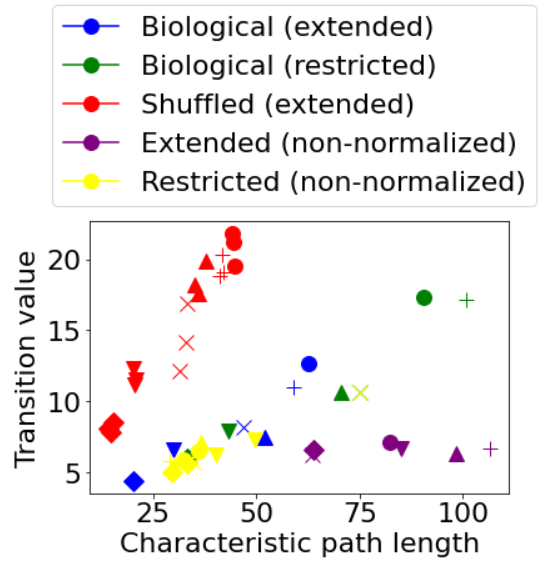}
    \caption{The same as Fig. \ref{all}, but with the inclusion of data points for non-normalized networks, i.e., networks in which edge weights were not divided by the sum of the volumes of its nodes. }
\label{allnon}     
\end{figure}

Fig. \ref{allnon} presents the analysis that incorporates connectomes which are not normalized, i.e., the connectivities are not divided by the sum of the volumes of the nodes. It is clear that the yellow points (representing the results for the restricted connectomes) fall on the biological trendline, but the purple points (representing the results for the extended connectomes) do not. The anomalous behavior of the latter is attributable to the presence of exceedingly dominant nodes such as the brainstem, the effect of which is regularized when their great size is taken into account in the normalized connectomes. \\ \indent

\begin{figure}[h]
\centering
\includegraphics[width=0.47\textwidth]{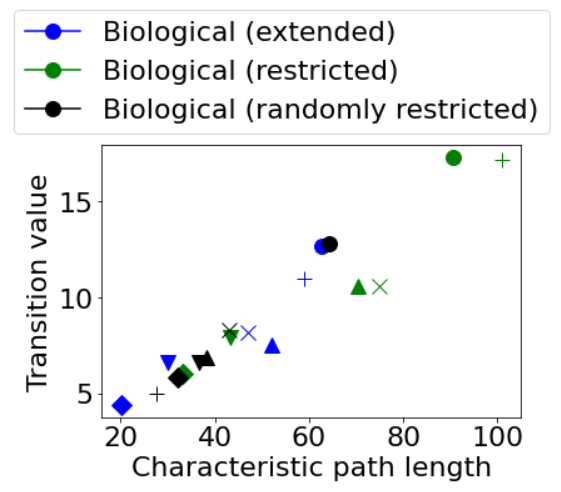}
    \caption{The transition value $c_5^T$ vs. the characteristic path length of a network. Symbol shapes correspond to subject names. Blue symbols are extended connectomes, green are restricted connectomes (in accordance with the FreeSurfer Desikan-Killiany atlas), and black are connectomes restricted by removing 20 randomly chosen regions (found in Table \ref{tab:removed_structures_2}. Errors are small than symbol sizes. }
\label{allrr}     
\end{figure}

\begin{figure}[h]
\centering
\includegraphics[width=0.47\textwidth]{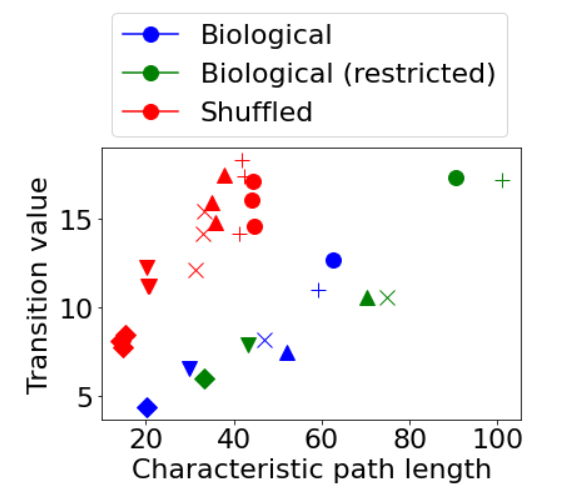}
    \caption{The same as Fig. \ref{all}, but with $c_5^T$ alternatively defined for the shuffled networks to be the lowest value of $c_5^T$ at which the system leaves the ground state, no matter how small the activity.}
\label{allalt}     
\end{figure}

Fig. \ref{allrr} compares the biological extended and restricted connectomes to another case of a restricted connectome in which the excluded structures are chosen randomly, rather than in line with the Desikan-Killiany atlas. One readily observes that while the green points (which correspond to the Desikan-Killiany atlas) are invariably further from the origin than their blue counterparts (corresponding to the extended connectomes) for each and every subject, the same is not true for the black points (corresponding to the randomly restricted connectomes) as compared with their blue counterparts. Indeed, for some subjects, the randomly restricted connectome is more well connected and more highly susceptible to global excitation, and for some other subjects, the converse is true. This suggests that the differences observed between the extended connectome and that which is restricted in accordance with the FreeSurfer Desikan-Killiany atlas are not purely caused by the lower system size of the latter, but rather by the exclusion of too important a set of structures. \\ \indent
Finally, Fig. \ref{allalt} presents the case in which we alternatively choose to define $c_5^T$ as the smallest value of $c_5$ at which the system departs from the ground state (no matter how meager the departure). A comparison with Fig. \ref{all} reveals that the only change is a slight lowering of the slope of the line of red points in Fig. \ref{allalt} with respect to those in Fig. \ref{all}, leaving our conclusions qualitatively unchanged.

\end{document}